# Architecture Optimization Dramatically Improves Reverse Bias Stability in Perovskite Solar Cells: A Role of Polymer Hole Transport Layers

Fangyuan Jiang,[1] Yangwei Shi,[1,2] Tanka R. Rana,[3] Daniel Morales,[4] Isaac Gould,[4] Declan P. McCarthy,[5] Joel Smith,[6] Grey Christoforo,[6] Hannah Contreras,[1] Stephen Barlow,[5] Aditya D. Mohite,[7] Henry Snaith,[6] Seth R. Marder,[4,5,8] J. Devin MacKenzie,[3,9] Michael D. McGehee,[4,5,8] David S. Ginger[1]*

1. Department of Chemistry, University of Washington, Seattle, WA 98195, USA

2. Molecular Engineering & Sciences Institute, University of Washington, Seattle, WA 98195, USA

3. Department of Materials Science and Engineering, University of Washington, Seattle, WA 98195, USA

4. Materials Science and Engineering Program, University of Colorado Boulder, Boulder, CO 80309, USA

5. Renewable and Sustainable Energy Institute, University of Colorado Boulder, Boulder, CO 80303, USA

6. Department of Physics, University of Oxford, Parks Road, Oxford OX1 3PU, UK

7. Department of Materials Science and NanoEngineering, Rice University, Houston, TX 77005, USA

8. Department of Chemical and Biological Engineering and Department of Chemistry, University of Colorado Boulder, Boulder, CO 80303, USA

9. Department of Mechanical Engineering, University of Washington, Seattle, WA 98195, USA

* Corresponding author: dginger@uw.edu

**Abstract**

We report that device architecture engineering has a substantial impact on the reverse bias instability that has been reported as a critical issue in commercializing perovskite solar cells. We demonstrate breakdown voltages exceeding -15 V in typical p-i-n structured perovskite solar cells via two steps: i) using polymer hole transporting materials; ii) using a more electrochemically stable gold electrode. While device degradation can be exacerbated by higher reverse bias and prolonged exposure, our as-fabricated perovskite solar cells completely recover their performance even after stressing at -7 V for 9 hours both in the dark and under partial illumination. Following these observations, we systematically discuss and compare the reverse bias driven degradation pathways in perovskite solar cells with different device architectures. Our model highlights the role of electrochemical reaction rates and species in dictating the reverse bias stability of perovskite solar cells.

**ToC graphic**

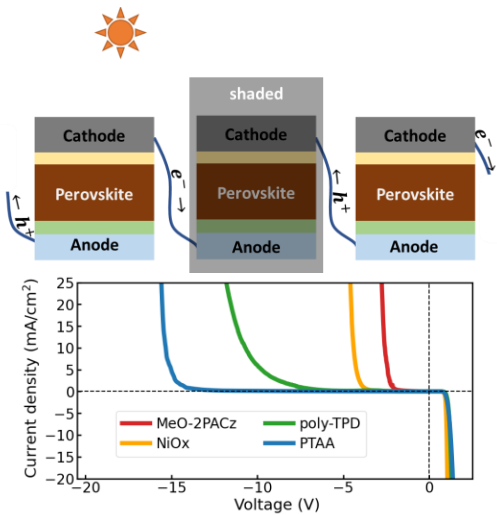

**Introduction**

Metal halide perovskite solar cells, with certified power conversion efficiencies (PCE) exceeding 26% (single junction) and 33% (perovskite-silicon tandem),[1] have emerged as competitive single-junction rivals to, and ideal tandem partners with the monocrystalline silicon solar cells that have dominated the photovoltaic (PV) market for decades. Early on, most efforts in perovskite PV were directed at making perovskite cells more efficient.[2–11] Recently, as the efficiency of perovskite PV made using low energy and low-capital-cost processes have begun to match and even exceed that of Si PV, more effort has been directed at issues of stability, with solar cells maintaining > 90% initial PCE after a few thousands of hours' aging at different ISOS (the International Summit on Organic Photovoltaic Stability) conditions,[12–14] and with various consortia emerging to develop testing protocol and recommendations for halide perovskite PV modules.[15]

Recently, a number of reports have highlighted reverse bias stability as a potentially critical issue facing perovskite PV, perhaps even the most demanding durability issue for fielded solar modules.[16–26] Reverse bias instability occurs when the low-output solar cells (e.g. due to partial shading) in serially connected modules are forced to match the current output of other high-output solar cells. The solar cells under reverse bias can also become quite hot, not only because the voltage is higher than under normal operation, but also because the breakdown current often passes through only part of the cell (causing localized heating). Such breakdowns are a big concern even for well-established silicon PV technologies. So far, relatively few reports have studied the reverse bias stability of perovskite solar cells. The breakdown voltages ($V_{RB}$s) in these studies typically range from -0.5 V to -3 V,[19–30] which still lag significantly behind typical Si $V_{RB}$ values of -13 V.[21,31] Increasing the $V_{RB}$ will reduce the number of bypass diodes needed to protect a panel and could make them much more cost-effective.[16,21] In specific, while one bypass diode is expected to protect ~21 silicon solar cells, the number drops to ~2 for perovskite solar cells considering their much lower $V_{RB}$.[21]

The lack of a mechanistic understanding of the reverse bias-driven breakdown process is a major challenge towards addressing the reverse bias stability issue in perovskite solar cells. The mainstream/predominant models gaining attention in the field today suggest that, under reverse bias, p-i-n structured perovskite solar cells breakdown commences at the perovskite/electron-transporting layer (ETL)/metal electrode side where oxidation reactions take place due to hole injection/tunneling.[19,20,23,32] These reactions range from the oxidation of the electrochemically active Ag electrode (which forms readily oxidized silver halides), to direct oxidation of iodide ions to form mobile $I_i^0$, $I_3^-$, $I_i^+$, or even release of $I_2$ gas.[23,30,32–38]

These models indeed explain many features observed to date, including the presence of Ag in the perovskite lattice following reverse bias,[19,32] the improvement in reverse bias stability when using less electrochemically active carbon as the electrode,[22] and when tailoring the interface between the perovskite and the ETL.[23] For instance, Huang and co-authors introduced a wide-bandgap lead sulfate layer or an insulating DPSI layer on top of perovskite with the intent of increasing the hole injection barrier from the $C_{60}$ side: they managed to stabilize perovskite solar cells at -1 V for 10 min with only < 20% efficiency loss.[23] While these are encouraging advances, silicon solar cells exhibit much more robust $V_{RB}$ values, typically as good as -13 V.[21,31]

Here, we systematically investigate the reverse bias behavior of typical p-i-n structured perovskite solar cells. We first try surface treatments to decrease halide vacancy concentrations at the perovskite/ETL interface and try inserting an alternative organic ETL between perovskite and $C_{60}$. We find neither approach yields visible $V_{RB}$ improvements. Next, we show that using thicker conjugated polymer films as the hole transporting layers (HTL) can lead to dramatic increases in reverse bias stability. Even with electrochemically-active Ag electrodes, using a thicker polymer HTL can improve the $V_{RB}$ to better than -8 V, as compared to worse than -1 V for phosphonic-acid-modified ITO interfaces that are in wide use today. Furthermore, replacing Ag with more electrochemically stable Au, we demonstrate a record $V_{RB}$ of < -15 V for perovskite solar cells, which is comparable to that for silicon solar cells and is much better than for other solar cell technologies.[17,31] Based on these experimental observations, we revisit previous proposals of reverse-bias-driven degradation mechanisms, and propose that planarization of the ITO electrode and blocking of electron injection under reverse bias need to be considered as important parts of the complete picture of reverse bias breakdown behavior.

**Results and Discussions**

We start from reverse bias stability study in an archetypal p-i-n structured perovskite solar cell shown in Figure 1a. We use a mixed-cation, triple-halide perovskite composition with the formula of $Cs_{0.22}FA_{0.78}Pb(I_{0.85}Br_{0.15})_3$ and with 3 mol % of $MAPbCl_3$ additive, where FA refers to formamidinium and MA refers to methylammonium. The perovskite composition, denoted as Cs22Br15Cl hereafter, has a relatively wide bandgap of 1.67 eV that is suitable for building perovskite-silicon tandems, and was previously reported to possess relatively good stability.[39] We choose a [2-(3,6-dimethoxy-9H-carbazol-9-yl)ethyl]phosphonic acid (MeO-2PACz) self-assembling monolayer (SAM) as the reference hole transporting interface modifier and archetypal evaporated $C_{60}$ as the ETL because they are commonly used and have been demonstrated to show good device performance in many previous reports.[40–43] We also employ Ag as a representative widely-used back (top) electrode. In the reverse-bias regime, we ground the Ag electrode, and apply a negative bias to the indium tin oxide (ITO) electrode. In this reverse-biased case, electrons will be injected from the cathodic ITO electrode while holes are being injected from the anodic Ag electrode.

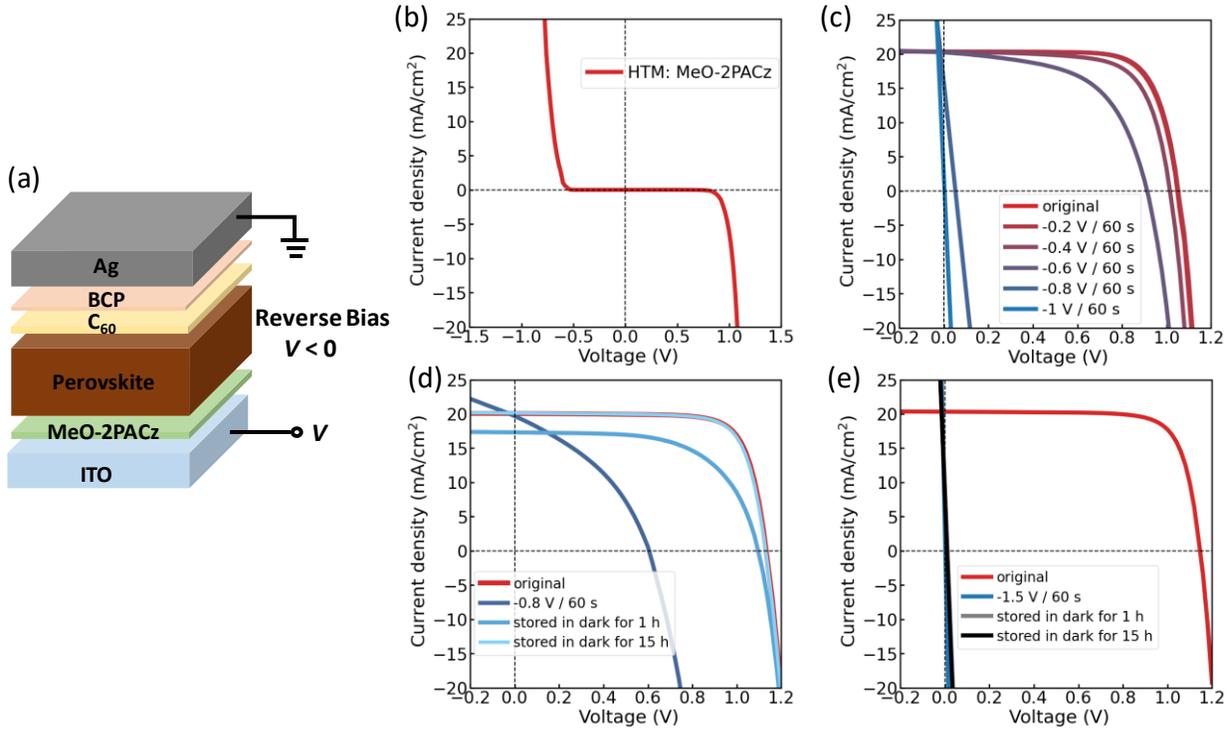

Figure 1. (a) Schematic diagram of an archetypal p-i-n structured perovskite solar cell under reverse bias. (b) Dark $J$-$V$ curves of perovskite solar cells with structure shown in (a). (c) $J$-$V$ curves (reverse scans) of the perovskite solar cell after holding at the stated reverse bias (gradually increased) for 60 s. (d) $J$-$V$ curves (reverse scans) showing that after biasing at -0.8 V for 60 s, device performance can be completely restored after 15 h dark storage. (e) $J$-$V$ curves (reverse scans) showing that device performance cannot be recovered after aging at -1.5 V for 60 s.

Figure S2(b) and Figure S3 show the current density – voltage ($J$-$V$) scans and statistics of the as-fabricated solar cells under forward and reverse scans. The best power conversion efficiencies (PCE) are 18.2% for both reverse and forward scan, consistent with expectations for this device stack.[44] We then investigate the reverse-bias behavior of these reference solar cells by measuring their dark $J$-$V$ curves (Figure 1b). We identify the breakdown voltage $V_{RB}$, defined as the onset voltage where the reverse current starts to increase dramatically (Figure S4), to be around -0.7 V. To monitor the reverse bias-induced degradation, we adopt a stress test in which we progressively increase the reverse bias in increments of -0.2 V at an interval of ~ 2 min. We note that, throughout the manuscript, we always refer to an increase in reverse bias as the increase in its absolute value (more negative), with low reverse bias meaning that the absolute value of the bias is low. After each step, we measure *J-V* curves under simulated 1-sun illumination to monitor the changes in device performance. Figure 1c displays the representative evolution of the *J-V* curves (reverse scans). As the reverse bias increases, the solar cell degrades further until it undergoes complete performance loss after being reversed biased at ~ -1 V. Consistent with previous studies,[19] we find that the solar cell performance loss during this reverse bias stressing undergoes two stages: first if the reverse bias is low (smaller than ~ -0.8 V), the solar cell can fully recover its performance after the removal of the bias, as shown in Figure 1d. On the contrary, if the reverse bias is high (more negative than ~ -1 V for this device architecture), the solar cell will be irreversibly shunted and its performance cannot be restored, as shown in Figure 1e.

Previous reports have identified the oxidation of the Ag electrode,[32,39] and the interaction between iodide and injected holes,[23,32] as major contributors to device degradation under reverse bias. Bertoluzzi et al. conducted drift-diffusion simulations to show that holes could tunnel into the perovskite and subsequently react with iodide interstitials, due to sharp band bending near the perovskite/electron-transporting contact.[30,45] These factors all point to the key role of the interfaces at the perovskite/ETL/metal electrode side. Therefore, we first focused our attempts on optimizing the interface between perovskite and $C_{60}$. Among various strategies, we attempted passivating the perovskite surface with (3-aminopropyl)trimethoxysilane (APTMS), which has been proven to reduce surface recombination velocities at both neat perovskite[46] and perovskite/fullerene interfaces,[43] as well as suppress bias-induced shifts in the surface potential.[47] All these behaviors suggest that APTMS can reduce halide vacancies and vacancy-mediated halide transport at the surface, which should, in theory, suppress the ionic conductivity of the perovskite, and the ability of halide ions to diffuse and react at the interface.[30] In addition, we attempted inserting an additional non-fullerene ETL to block possible ion diffusion or charge injection, as well as slow down possible reactions of the halide with the fullerene.[48,49] Figure S5 compares the typical *J-V* curves and reverse bias behaviors of reference and APTMS-passivated solar cells. While APTMS surface treatment improves the $V_{OC}$ from 1.10 V to 1.14 V, consistent with previous reports,[43,46] we did not observe any improvement in the reverse bias behavior. Similarly, Figure S6 shows that the additional electron transporting layer, a naphthalene diimide derived polymer named NDI-1,[50] also has no beneficial impact on the reverse bias behavior. These experimental observations led us to consider whether there might be other pathways to improve $V_{RB}$.

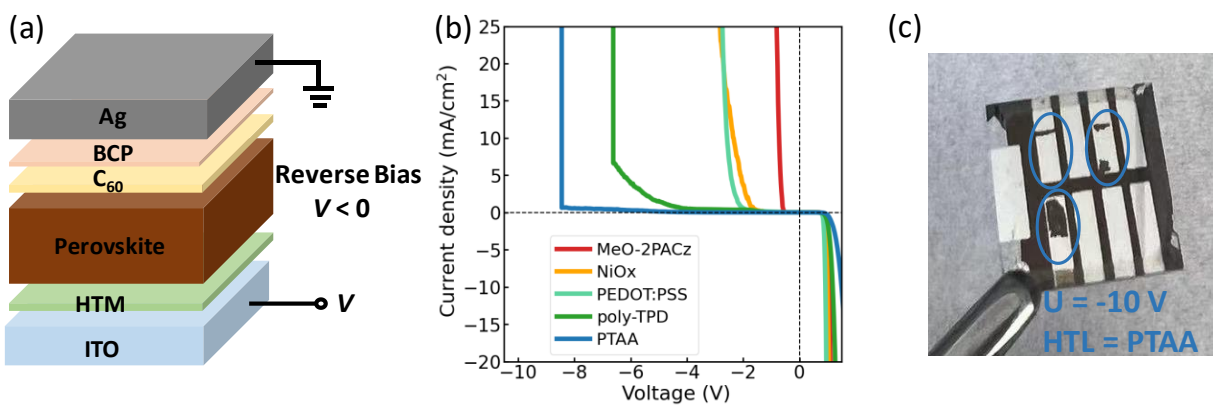

Figure 2. (a) Schematic diagram of a typical p-i-n structured perovskite solar cell under reverse bias. (b) Dark *J-V* curves of perovskite solar cells with different hole transporting materials. (c) Image of PTAA-based perovskite solar cells after dark *J-V* sweeping to -10 V.

We next explore whether replacing the commonly used phosphonic acid HTL could improve reverse bias performance. This choice may seem surprising since, according to the mainstream models for explaining reverse bias breakdown,[23,30] changing the HTL is not expected to have a significant effect on $V_{RB}$. We selected a few HTLs (molecular structures shown in Figure S1), including the reference phosphonic acid MeO-2PACz, NiO$_x$, poly(3,4-ethylenedioxythiophene) polystyrene sulfonate (PEDOT:PSS), poly(4-butyl-triphenylamine-4',4"-diyl) (poly-TPD), and poly(2,4,6-trimethyltriphenylamine-4',4"-diyl) (PTAA). Our intention was to compare how these HTLs, including a metal oxide (NiO$_x$), a small molecule self-assembled monolayer (MeO-2PACz), and different polymers (PEDOT:PSS, poly-TPD, PTAA) affect the reverse bias behavior of perovskite solar cells if all other layers are the same. We first optimized the processing condition of these HTLs to ensure good device performance, as shown in Figure S2 and Figure S3. We also optimized the processing of perovskite films on top of these HTLs to ensure good crystallinity and morphology, as characterized by x-ray diffraction (XRD, Figure S7) and scanning electron microscopy (SEM, Figure S8). We decided on thicknesses of MeO-2PACz, NiO$_x$, PEDOT:PSS, poly-TPD, PTAA as < 5 nm (presumably monolayer[51,52]), 40 nm, 29 nm, 16 nm, and 35 nm respectively. Besides, we introduced an additional PFN-Br layer (< 5 nm) on the top surface of PTAA and poly-TPD to increase the wettability of the perovskite solution.

Figure 2b shows these HTLs do lead to significantly different reverse bias behaviors, with $V_{RB}$ values of -0.7 V for MeO-2PACz, -1.9 V for NiO$_x$, -2.4 V for PEDOT:PSS, -6.6 V for poly-TPD, and -8.4 V for PTAA-based perovskite solar cells (control experiments with only PFN-Br interlayer are shown in Figure S9, which rules out the possibility that it is the PFN-Br that caused the $V_{RB}$s difference). We see that the

MeO-2PACz SAM exhibits a much weaker reverse bias stability than other counterparts, especially compared to polymers (PEDOT:PSS, PTAA, poly-TPD). Meanwhile, we note that our results are consistent with previous report by Razera and coauthors where improved reverse bias stability are associated with perovskite solar cells using a $NiO_x$ HTL.[19] Our results are also consistent with perovskite photodetector work reported by Li[53] and Tsai[54], where improved $V_{RB}$s of up to -7.5 V were achieved in perovskite photodetectors which both coincidentally used PTAA as the HTL. Even though our results differ from these photodetector studies in terms of device structures, materials, and strategies, having a robust PTAA interlayer seems to be a common characteristic to achieve high stability under high reverse bias.

For a PTAA-based solar cell, as we further increase the bias to exceed the $V_{RB}$ during the *J-V* sweeping of solar cells, we observe irreversible damage to the Ag electrode that is sufficiently severe to be visible by the naked eye. Figure 2c and Figure S10 show that these catastrophic failures are also accompanied by large currents flowing through the cell.

Ag is relatively easy to oxidize, especially in the presence of halide species,[55] and is a fairly soft metal with relatively mobile cations. Therefore, to test if the electrochemistry of Ag[32] limits reverse bias stability, we next fabricated devices with identical devices (Figure 3a) comparing Ag and Au as the back metal. Figure 3b clearly demonstrates that replacing Ag with the more electrochemically stable Au yields even greater improvements in reverse bias stability. We measured the dark *J-V* curves and found the $V_{RB}$s to be -2.5 V for MeO-2PACz, -4.3 V for $NiO_x$, -10.2 V for poly-TPD, and -15.3 V for PTAA-based perovskite solar cells. This trend is consistent with that of Ag-based solar cells, yet with larger overall breakdown voltages. The $V_{RB}$ increase from -2.5 V (MeO-2PACz) to -15.3 V (PTAA) again verifies that the choice of HTL can significantly improve the cell robustness to reverse bias breakdown. Notably, the $V_{RB}$ < -15 V achieved in PTAA-based perovskite solar cell, which is reproducible across multiple devices and batches (Figure S11), is comparable to, or even better than values reported for commercial PV technologies, such as Si, CIGS, and CdTe.[17,21,31] Figure 3c compares the reverse current in log scale among different perovskite solar cells. The PTAA-based devices show much lower (~ a few orders of magnitude) reverse-bias dark current than that of other non-PTAA based solar cells. Importantly, we note that the $V_{RB}$s for all the Au-based perovskite solar cells are better than that of the Ag-based cell made with the same HTLs, suggesting that the $V_{RB}$ may depend on interactions spanning both interfaces. We explore this observation further below.

We also explore the reverse bias stability of perovskite solar cells when illuminated with dim light (< 0.1 Sun), as shown in Figure S12-13. Such illumination serves to simulate the real outdoor environment for solar modules when the low power-output cells are not shaded under complete darkness. We find that, in the presence of < 0.1 Sun illumination, the $V_{RB}$s, are -2.4 V with MeO-2PACz (Figure S12(c)) and ~ -15 V with PTAA (Figure S13(c)), remaining similar with that in complete darkness.

We note that voltage-induced solar cell degradation is not only related to the magnitude of the reverse bias, but also the duration of the applied reverse bias. Thus, to show that the improved stability with thicker polymer HTLs is not an artifact of scan/sweep rates (Figure S14), we study the reverse bias behavior of PTAA-based perovskite solar cells over an extended duration. Figure 3d tracks the reverse current densities of perovskite solar cells aged at -7 V for 9 h in the dark and under dim light (~ 0.1 Sun illumination). We can see that the current densities remain low despite a slight increase over time. In contrast, as we increase the bias to -10 V (Figure 3e), the reverse current density initially remains low/stable, but dramatically increases after around 2 h which can be ascribed to device shunting.

We also conduct *J-V* measurements to monitor the performance changes of PTAA-based solar cells after aging at various bias for different durations. While no device degradation occurs when solar cells are stressed at -7 V for a relatively short duration of 1800 s (30 min), (Figure S15a), extending the duration to 9 h leads to slight device performance loss (Figure 3f and Figure S15(b)). Notably the performance of the devices is completely recoverable even under such lengthy exposure to such conditions. As we increase the bias to -10 V, the perovskite solar cell starts to show obvious degradation after aging for 10 min, which, fortunately, is mostly recoverable after only 10 min under illumination and entirely recoverable after dark storage, as shown in Figure S15(c).

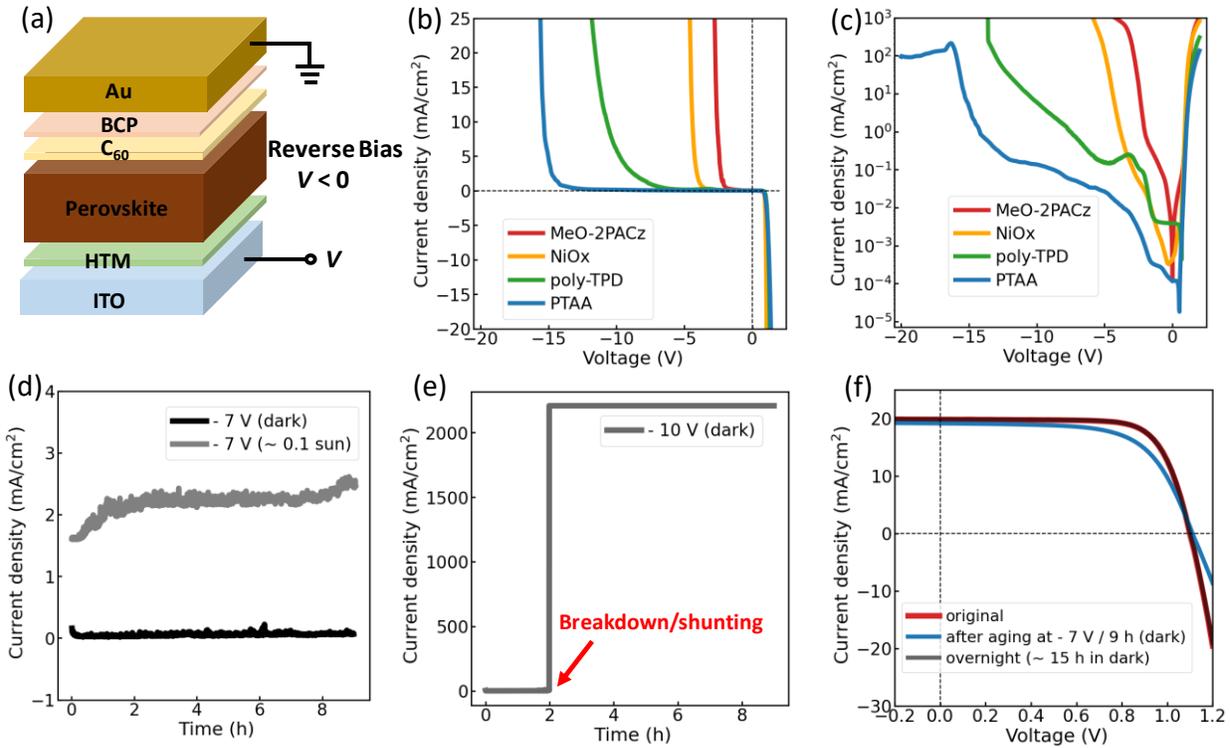

Figure 3. (a) Schematic diagram of the modified p-i-n structured perovskite solar cell employing Au as the top electrode, under reverse bias. Dark *J-V* curves in (b) linear and (c) log scale of perovskite solar cells with different HTLs. Current density tracking of PTAA-based perovskite solar cells stressed (3d) at -7 V for 9 h in the dark or under ~0.1 Sun dim light illumination, and (3e) at -10 V for 9 h in the dark where solar cell breakdown takes place at around ~2 h. (f) *J-V* curves (reverse scans) showing that device performance can be restored after aging at -7 V for 9 h.

From these observations, we conclude that the choice of HTLs and metal electrode matters a great deal to reverse bias stability. Given the previous emphasis on the perovskite/electron transport layer interface[23,30,32] and the strong experimental evidence that oxidation reactions are occurring at that interface under reverse bias,[19] the dramatic influence of the HTLs on reverse bias stability may seem surprising. Thus, we consider why the choice of polymer HTLs could make such a large difference. We examine four main hypotheses: (1) the polymer HTLs could somehow store reactive halogen species,[56] therefore delaying the onset of permanent performance loss; (2) the polymer HTLs could somehow drop a substantial fraction of the reverse voltage, reducing the field experienced by the perovskite and thus indirectly suppressing hole injection at the anodically-biased ETL interface; (3) the role of electron injection under reverse bias has been always overlooked, and these HTLs provide different levels of electron blocking, and/or (4) the possibility that the thick polymer HTLs provides a physical barrier to reduce shorting/shunting at imperfections in the device.

Ultimately, we dismiss the first two hypotheses in favor of the third and fourth (Figure S16 – S18, with detailed calculations and analyses). First, when considering the role of the polymer HTLs as a possible halide "sponge," we note that even a "deep-HOMO" polymer like poly-vinylecarbazole (PVK) still improves the reverse bias stability significantly (Figure S16). This observation effectively rules out iodine/polymer reactions as the source of improved reverse bias stability. Second, we examine the effects of the thicker HTLs on the electric field distribution in the device, using both back-of-the-envelope calculation (Figure S17), full drift-diffusion simulation (Figure S18), as well as analysis of the device *J-V* curves (Figure S2 – 3). While there are a wide range of variables, we conclude that it is unlikely for the addition of even a "thick" polymer HTL (~30 nm) into a ~600-nm-thick device stack to lead to the necessary changes in field profiles in the perovskite. And that if there was a substantial voltage drop across the HTL, the devices would have a remarkably poor fill factor.

We now discuss the possibility that the electron injection under reverse bias, which has been always overlooked by the field, is playing an essential role in mitigating reverse bias breakdown of perovskite solar cells. From previously established evidence, both oxidation of the metal electrode and oxidation of iodide species seem likely to take place.[19,23,30,32] Nevertheless, *charge conservation requires that oxidation and reduction reactions occur in pairs*.[57] From a device physics perspective, oxidation without a corresponding reduction corresponds to injection of uncompensated positive space charges, which would lead to a large field that would oppose the further hole injection. From an electrochemical perspective, these positive charges would correspond to an increase in the oxidation potential, slowing down further oxidation.

We estimated such an effect based on the Poisson equation (see SI for detailed calculations and analyses): an injected hole (oxidized species) density of ~1.4 x $10^{16}$ cm$^{-3}$, without corresponding electrons (reduction reactions), would be sufficient to completely counter an applied bias of 10 V across a 460-nm-thick perovskite film. Such a hole density corresponds to oxidation of < 1% of a monolayer of Ag atoms at metal/ETL interface or the oxidation of 0.00015% of the iodide in the perovskite. With these factors in mind, we propose a modification to the current picture of oxidation-driven degradation of p-i-n structured perovskite solar cell devices under reverse bias: we speculate that the metal electrode and iodide oxidation events do take place as previously reported,[19,23,30,32] *however, the progression of these oxidations must be paired with reduction/electron injection events at the cathodically-biased HTL/perovskite interface*. In this case, the ability of the HTLs to affect electron injection (slow down reduction reactions) would have a direct impact on the oxidation reactions (at the perovskite/ETL/metal side) by preventing oxidation from progressing substantially when a paired reduction/electron injection event is missing.

We also consider that the role of PTAA in improving reverse bias stability may arise from its ability, given its polymeric nature and relatively thick film, to better homogenize/planarize the bottom transparent conductive oxide (TCO) contact. In the organic light-emitting diode (OLED) fields, it is well known that planarization of the TCO is essential for device stability and yield.[58–63] We believe that the thick polymer HTLs are playing similar roles in perovskite PV, especially under reverse bias where the field and/or current would be concentrated on any imperfections, defects, pinholes, or dust particles in the device and drive localized electrochemical breakdown. We note that this mechanism is therefore not distinct from the role of HTLs in blocking electron injection and preventing electrochemical breakdown pathways, and that thick polymer HTLs would synergistically provide both advantages.

We now discuss what the reduction reactions at the ITO/ HTL/perovskite side could be. We speculate that the reduction of $Pb_i^{2+}$ (shadow defects) into $Pb_i^0$ (deep defects) by injected electrons is most likely. This step is plausible as the reduction of $Pb^{2+}$ is electrochemically facile compared to halide oxidation or even reduction of silver halides salts.[55] Furthermore, previous XPS measurements have confirmed that $Pb^0$ impurities usually exist in perovskite as a decomposition byproduct, and these $Pb^0$ impurities are generally associated with degradation of solar cell performance.[64–66] This step is also consistent with the observation of PL quenching fronts propagating from the negatively-biased electrode in coinage-metal/perovskite junctions.[67] Most importantly, such PL quenching has been proven to be reversible,[67] which is consistent with the observed recoverable device performance loss in our perovskite solar cells under low reverse bias regime (Figure 1d).

In Figure 4, we propose a set of dominant electrochemical reactions (oxidation – reduction pairs) that may occur under reverse bias in perovskite solar cells with Ag or Au electrodes. To better understand the degradation process, a more detailed schematic can be found in Figure S19 along with more analysis explaining how these reactions take place in perovskite solar cells with different architectures. We note that the perovskite degradation processes are very complex, thus we do not anticipate our model to be the only/ultimate mechanism. However, we put this framework forward as a self-consistent proposal that is supported by our experimental data, as well as with various analytical studies in the literature.[19,23,32]

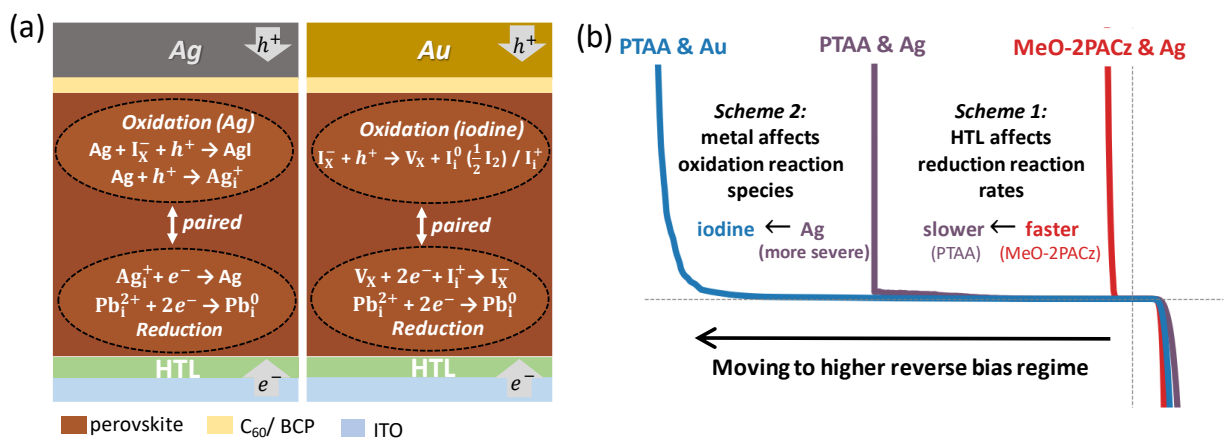

Figure 4. Schematic of degradation mechanisms of perovskite solar cells under reverse bias. (a) Paired electrochemical reactions in perovskite solar cells with Ag or Au top electrode; (b) Schematic suggesting that the role of PTAA HTLs is to slow down reduction reaction rate (thus electrochemical degradation), while the role of Au electrode is to replace Ag oxidation with less severe/deadly iodine oxidation, both strategies contributed to the record $V_{RB}$ in this study.

In a perovskite solar cell with Ag electrode (Figure 4a, left), consistent with previous reports,[19,32] we believe that Ag oxidation will first take place by injected holes to form AgI and $Ag_i^+$, given the relatively low oxidation potential of Ag in the presence of iodide.[55] However, Ag oxidation reactions cannot proceed if in the absence of necessarily paired reduction reactions. In other words, perovskite solar cell device degradation, which is directly linked to electrochemical reactions, relies on the reduction reaction (electron injection) rates at the HTL/perovskite interface. Since different HTLs provide different levels of electron blocking, and since they also serve as different levels of physical barriers to reduce shorting/shunting at imperfections in the device, a good HTL is therefore expected to slow down the reduction reaction rate (due to suppressed electron injections) and therefore slow down the necessarily paired Ag oxidations at the anodically biased contact.

Compared with MeO-2PACz SAMs (< 5 nm), which allow electrons to transport/tunnel through easily, thick polymer PTAA layers (~35 nm) offer better electron blocking ability, therefore pushing the electrochemical degradation of perovskite solar cells to higher reverse bias regime, shown as *Scheme* 1 in Figure 4b. Once the electron injections (reduction reactions) occur at the ITO/HTL side, the progression of Ag oxidation reactions becomes possible. In this case, more Ag will get oxidized to form AgI and $Ag_i^+$, with the latter diffusing and/or drifting towards the cathodically-biased ITO side under the electrical field, until being reduced by the injected electrons to form metallic Ag. As the reverse bias increases with more injected carriers, these metallic Ag will form conductive Ag filaments within the perovskite layer in a manner analogous to that proposed for memristive devices (Equation 5).[68–70] We note that the formation of Ag filaments in perovskite layer following reverse bias stressing has been widely observed and characterized.[19,32] It is the essential role of a HTL in dictating the electron injection (electrochemical reaction pairs) and thus the formation of Ag filaments that has always been overlooked by the field.

After replacing Ag with Au (Figure 4a, right), which is more electrochemically stable and possesses a much higher oxidation potential,[55] since both Au and $Br_X^-$ have higher oxidation potentials than $I_X^-$,[55] $I_X^-$ tends to be oxidized first, producing oxidized iodine species, such as $I_i^0$ and $I_i^+$.[23,32] The $I_i^+$ species can diffuse and/or drift within the perovskite layer to the cathodically-biased ITO side under electrical field, until they are electrochemically reduced into $I_X^-$ by injected electrons. This process has been experimentally verified in an analytical study by Xu and coauthors via ToF-SIMS measurement where they observed negligible Au signal within the solar cell device after reverse bias stressing. They also observed halide segregation with bromide and iodide accumulating at the Au and ITO side respectively via PL and EDX measurements, which is a result of $I_X^-/I_i^+$ redox shuttle and subsequent halide transport process that is indicative of iodine oxidation.[32] It is notable that the $I_i^0$ could form $I_2$ which may escape from the perovskite to cause irreversible device performance loss.[30]

We also note that there might be other electrochemical reactions (degradation pathways) taking place within the solar cell, especially as the reverse bias increases. For example, with many injected carriers, the perovskite might suffer from significant loss of iodide, resulting in the formation of metallic Pb, which later might form Pb filaments that shunt the solar cell device.[19] Also, the charge transporting layers might suffer from dielectric breakdown under high reverse bias. We speculate that, in terms of solar cell performance loss under reverse bias, neither the degradation of the perovskite film nor the interlayers (in the case of Au electrode) are as severe/deadly as the formation of Ag filaments (in the case of Ag electrode). Therefore, replacing Ag with more electrochemical-stable Au allows substituting Ag oxidation with more benign iodine oxidation, and thus extending the electrochemical degradation of the cells to higher reverse bias regime, shown as *Scheme* 2 in Figure 4b. Although Au is likely undesirable for a scaled device, there are a number of alternative electrode materials for the electron-extracting electrode that should be superior to Ag, such as carbon and TCOs.

**Conclusions**

In summary, we systematically investigated factors that could affect the reverse bias behavior of a series of p-i-n type perovskite solar cells, ranging from passivation of halide vacancies and inserting an additional ETL between the perovskite and the fullerene, to systematically varying the HTLs and metal electrodes. Our results suggest that engineering the HTLs and metal electrodes is of vital importance for preventing breakdown of p-i-n structured perovskite solar cells. Even in the presence of a more "active" Ag electrode, using a robust electron-blocking PTAA HTL improves the $V_{RB}$ to < -8 V, as compared with that of > -1 V for MeO-2PACz based perovskite solar cells. Further optimization on the metal electrode via replacing Ag with Au extends the $V_{RB}$ to < -15 V. Our results suggest that efforts to achieve high reverse bias stability in manufactured perovskite p-i-n devices should focus on (1) using electrodes that are more electrochemically inert (not just Au, but carbon or TCOs); (2) using interlayers that can more robustly block hole and/or electron injection under reverse bias; (3) optimizing the processing of each functioning layers to avoid pinholes or otherwise defective sites that might cause leakage or shunting of the device. Using these strategies, the optimization of the perovskite device architecture could allow realization of large area p-i-n perovskite solar cells with $V_{RB}$ comparable to Si photovoltaic technologies.

**Acknowledgments**

This work, and the roles of F.J., Y. S, D.P.M, J. S., G.C., S.B., S.R.M, H. S., and D.S.G. were primarily supported by the Office of Naval Research (Award # N00014-20-1-2587). F. J. and D.S.G. acknowledge the institutional support from the B. Seymour Rabinovitch Endowment and the state of Washington. The authors acknowledge the use of facilities and instruments at the Photonics Research Center (PRC) at the Department of Chemistry, University of Washington, as well as that at the Research Training Testbed (RTT), part of the Washington Clean Energy Testbeds system. Part of this work was carried out at the Molecular Analysis Facility, a National Nanotechnology Coordinated Infrastructure site at the University of Washington which is supported in part by the National Science Foundation (NNCI-1542101), the Molecular Engineering & Sciences Institute, and the Clean Energy Institute. I.G and M.D.M acknowledge support by the U.S. Department of Energy's Office of Energy Efficiency and Renewable Energy (EERE) under Solar Energy Technologies Office (SETO) Agreement Number DE-EE0009513. T. R. R. and J. D. M. acknowledge support from the Washington Research Foundation, the University of Washington Clean Energy Institute's Washington Clean Energy Testbeds, and the Department of Energy's SETO through the Perovskite Photovoltaic Accelerator for Commercializing Technologies program. F. J. especially acknowledges all scientific discussions with Jiajie Guo, Muammer Yaman, and Shinya E. Chen, all of the whom are graduate students at University of Washington.

**Author contributions**

F.J. and D.S.G. conceived the project, designed the experiments, and discussed the results together. F. J., Y. S., and H. C. performed the experiments and analyzed the data. T. R. R. and J. D. M. provided the NiOx. D. P. M., S. B., and S. R. M. provided the NDI-1 electron transporting material. D. M., I. G., M. D. M., and A. D. M. contributed to the electric field screening calculation and discussions. J. S., G. C. and H. S. helped with the standardization of J-V characterizations and definition of breakdown voltage. All authors contributed to the interpretation of the data as well as the presentation of this manuscript. F. J. wrote the first draft. F. J and D.S.G. revised the manuscript with input from all the authors.

**Competing interests**

Michael D. McGehee is an advisor to Swift Solar. Henry Snaith is a co-founder of Oxford PV. The other authors declare no competing interests.

**References**

1. Best Research-Cell Efficiency Chart (accessed on August 2023), https://www.nrel.gov/pv/cell-efficiency.html.

2. Lee, M. M. et al. Efficient Hybrid Solar Cells Based on Meso-superstructured Organometal Halide Perovskites. *Science* **338**, 643–647 (2012).

3. Jeon, N. J. et al. Solvent Engineering for High-Performance Inorganic-organic Hybrid Perovskite Solar Cells. *Nat. Mater.* **13**, 897–903 (2014).

4. Yang, W. S. et al. Iodide Management in Formamidinium-lead-halide-based Perovskite Layers for Efficient Solar Cells. *Science* **356**, 1376-1379 (2017).

5. Lu, H. et al. Vapor-Assisted Deposition of Highly Efficient, Stable Black-phase $FAPbI_3$ Perovskite Solar Cells. *Science* **370**, eabb8985 (2020).

6. Jeon, N. J. et al. Compositional Engineering of Perovskite Materials for High-performance Solar Cells. *Nature* **517**, 476–480 (2015).

7. Yang, W. S. et al. High-performance Photovoltaic Perovskite Layers Fabricated through Intramolecular Exchange. *Science* **348**, 1234–1237 (2015).

8. Li, X. et al. A Vacuum Flash-assisted Solution Process for High-efficiency Large-area Perovskite Solar Cells. *Science* **353**, 58–62 (2016).

9. Zhou, H. et al. Interface Engineering of Highly Efficient Perovskite Solar Cells. *Science* **345**, 542–546 (2014).

10. Jeong, J. et al. Pseudo-halide Anion Engineering for A-$FAPbI_3$ Perovskite Solar Cells. *Nature* **592**, 381–385 (2021).

11. Li, N. et al. Liquid Medium Annealing for Fabricating Durable Perovskite Solar Cells with Improved Reproducibility. *Science* **373**, 561–567 (2021).

12. Khenkin, M. V. et al. Consensus Statement for Stability Assessment and Reporting for Perovskite Photovoltaics Based on ISOS Procedures. *Nat. Energy* **5**, 35–49 (2020).

13. Li, C. et al. Rational Design of Lewis Base Molecules for Stable and Efficient Inverted Perovskite Solar Cells. *Science* **379**, 690-694 (2023).

14. Azmi, R. et al. Damp Heat-stable Perovskite Solar Cells with Tailored-dimensionality 2D/3D Heterojunctions. *Science* **376**, 73-77 (2022).

15. Perovskite PV Accelerator for Commercializing Technologies (PACT), https://pvpact.sandia.gov/.

16. Lan, D. et al. Combatting Temperature and Reverse-bias Challenges Facing Perovskite Solar Cells. *Joule* **6**, 1782-1797 (2022).

17. Wang, C. et al. Perovskite Solar Cells in the Shadow: Understanding the Mechanism of Reverse-bias Behavior towards Suppressed Reverse-bias Breakdown and Reverse-bias Induced Degradation. *Adv. Energy Mater.* **13**, 2203596 (2023).

18. Boyd, C. C. et al. Understanding Degradation Mechanisms and Improving Stability of Perovskite Photovoltaics. *Chem. Rev.* **119**, 3418–3451 (2019).

19. Razera, R. A. Z. et al. Instability of p-i-n Perovskite Solar Cells under Reverse Bias. *J. Mater. Chem. A*, **8**, 242–250 (2020).

20. Bowring, A. R., et al. Reverse Bias Behavior of Halide Perovskite Solar Cells. *Adv. Energy Mater.* **8**, 1702365, (2018).

21. Wolf, E. J. et al. Designing Modules to Prevent Reverse Bias Degradation in Perovskite Solar Cells when Partial Shading Occurs. *Solar RRL* **6**, 2100239 (2022).

22. Bogachuk, D. et al. Perovskite Photovoltaic Devices with Carbon-Based Electrodes Withstanding Reverse-Bias Voltages up to –9 V and Surpassing IEC 61215:2016 International Standard. *Solar RRL* **6**, 2100527 (2022).

23. Ni, Z. et al. Evolution of Defects during the Degradation of Metal Halide Perovskite Solar Cells under Reverse Bias and Illumination. *Nat. Energy* **7**, 65–73 (2022).

24. Najafi, L. et al. Reverse-Bias and Temperature Behaviors of Perovskite Solar Cells at Extended Voltage Range. *ACS Appl. Energy Mater.* **5**, 1378–1384 (2022).

25. Li, W. et al. Sparkling Hot Spots in Perovskite Solar Cells under Reverse Bias. *ChemPhysMater* **1**, 71–76 (2022).

26. Qian, J. et al. Destructive Reverse Bias Pinning in Perovskite/silicon Tandem Solar Modules Caused by Perovskite Hysteresis under Dynamic Shading. *Sustain. Energy Fuels* **4**, 4067–4075 (2020).

27. Ma, Y. et al. Suppressing Ion Migration across Perovskite Grain Boundaries by Polymer Additives. *Adv. Funct. Mater.* **31**, 2006802 (2021).

28. Jeangros, Q. et al. In-situ TEM Analysis of Organic-inorganic Metal-halide Perovskite Solar Cells under Electrical Bias. *Nano Lett.* **16**, 7013–7018 (2016).

29. Kim, D. et al. Light- and Bias-induced Structural Variations in Metal Halide Perovskites. *Nat. Commun.* **10**, 444 (2019).

30. Bertoluzzi, L. et al. Incorporating Electrochemical Halide Oxidation into Drift-diffusion Models to Explain Performance Losses in Perovskite Solar Cells under Prolonged Reverse Bias. *Adv. Energy Mater.* **11**, 2002614 (2021).

31. Breitenstein, O. et al. Understanding Junction Breakdown in Multicrystalline Solar Cells. *J. Appl. Phys.* **109**, 071101 (2011).

32. Xu, Z. et al. Halogen Redox Shuttle Explains Voltage-induced Halide Redistribution in Mixed-halide Perovskite Devices. *ACS Energy Lett.* **8**, 513–520 (2023).

33. Fu, F. et al. $I_2$ Vapor-Induced Degradation of Formamidinium Lead Iodide Based Perovskite Solar Cells under Heat-light Soaking Conditions. *Energy Environ. Sci.* **12**, 3074–3088 (2019).

34. Meggiolaro, D. et al. Iodine Chemistry Determines the Defect Tolerance of Lead-halide Perovskites. *Energy Environ. Sci.* **11**, 702–713 (2018).

35. Park, J. S. et al. Accumulation of Deep Traps at Grain Boundaries in Halide Perovskites. *ACS Energy Lett*. **4**, 1321–1327 (2019).

36. Motti, S. G. et al. Controlling Competing Photochemical Reactions Stabilizes Perovskite Solar Cells. *Nat. Photon.* **13**, 532–539 (2019).

37. Kerner, R. A. et al. The Role of Halide Oxidation in Perovskite Halide Phase Separation. *Joule* **5**, 2273–2295 (2021).

38. Motti, S. G. et al. Controlling Competing Photochemical Reactions Stabilizes Perovskite Solar Cells. *Nat. Photon.* **13**, 532–539 (2019).

39. Xu, J. et al. Triple-Halide Wide-band Gap Perovskites with Suppressed Phase Segregation for Efficient Tandems. *Science* **367**, 1097-1104 (2020).

40. Al-Ashouri, A. et al. Monolithic Perovskite/silicon Tandem Solar Cell with >29% Efficiency by Enhanced Hole Extraction. *Science* **370**, 1300-1309 (2020).

41. Jiang, Q. et al. Surface Reaction for Efficient and Stable Inverted Perovskite Solar Cells. *Nature* 611, 278–283 (2022).

42. Li, L. et al. Flexible All-perovskite Tandem Solar Cells Approaching 25% Efficiency with Molecule-bridged Hole-selective Contact. *Nat. Energy* **7**, 708–717 (2022).

43. Shi, Y. et al. (3-Aminopropyl)trimethoxysilane Surface Passivation Improves Perovskite Solar Cell Performance by Reducing Surface Recombination Velocity. *ACS Energy Lett*. **7**, 4081–4088 (2022).

44. Taddei, M. et al. Ethylenediamine Addition Improves Performance and Suppresses Phase Instabilities in Mixed-halide Perovskites. *ACS Energy Lett*. **7**, 4265–4273 (2022).

45. Bertoluzzi, L. et al. Mobile Ion Concentration Measurement and Open-access Band Diagram Simulation Platform for Halide Perovskite Solar Cells. *Joule* **4**, 109–127 (2020).

46. Jariwala, S. et al. Reducing Surface Recombination Velocity of Methylammonium-free Mixed-cation Mixed-halide Perovskites Via Surface Passivation. *Chem. Mater*. **33**, 5035–5044 (2021).

47. Pothoof, J. et al. Surface Passivation Suppresses Local Ion Motion in Halide Perovskites. *J. Phys Chem. Lett.* **14**, 6092–6098 (2023).

48. Guo, H. et al. Immobilizing Surface Halide in Perovskite Solar Cells via Calix[4]pyrrole. *Adv. Mater*. **35**, 2301871 (2023).

49. Boehm, A. M. et al. Influence of Surface Ligands on Energetics at $FASnI_3/C_{60}$ Interfaces and Their Impact on Photovoltaic Performance. *ACS Appl. Mater. Interfaces* **12**, 5209–5218 (2020).

50. Al Kurdi, K. et al. A Naphthalene Diimide Side-chain Polymer as an Electron-extraction Layer for Stable Perovskite Solar Cells. *Mater. Chem. Front*. **5**, 450–457 (2021).

51. Phung, N. et al. Enhanced Self-Assembled Monolayer Surface Coverage by ALD NiO in p-i-n Perovskite Solar Cells. *ACS Appl. Mater. Interfaces* **14**, 2166–2176 (2022).

52. Paniagua, S. A. et al. Phosphonic Acids for Interfacial Engineering of Transparent Conductive Oxides. *Chem. Rev*. **116**, 7117–7158 (2016).

53. Li, R. et al. Layered Perovskites Enhanced Perovskite Photodiodes. *J. Phys. Chem. Lett.* **12**, 1726–1733 (2021).

54. Tsai, H. et al. Addressing the Voltage Induced Instability Problem of Perovskite Semiconductor Detectors. *ACS Energy Lett.* **7**, 3871–3879 (2022).

55. Bard, A. J. et al. Standard Potentials in Aqueous Solution. (2017).

56. Xu, Z. et al. Origins of Photoluminescence Instabilities at Halide Perovskite/Organic Hole Transport Layer Interfaces. *J. Am. Chem. Soc.* **145**, 11846–11858 (2023).

57. Allen J. Bard, et al. Electrochemical Methods: Fundamentals and Applications. (2000).

58. Park, S.-J. et al. Enhancement of Light Extraction Efficiency of OLEDs Using $Si_3N_4$-based Optical Scattering Layer. *Opt. Express* **22**, 12392 (2014).

59. Sohn, S., et al. Printed Organic Light-Emitting Diodes on Fabric with Roll-to-Roll Sputtered ITO Anode and Poly(vinyl alcohol) Planarization Layer. *ACS Appl. Mater. Interfaces* **13**, 28521–28528 (2021).

60. Donie, Y. J. et al. Planarized and Compact Light Scattering Layers Based on Disordered Titania Nanopillars for Light Extraction in Organic Light Emitting Diodes. *Adv. Opt. Mater.* **9**, 2001610 (2021).

61. Dykstra, E. et al. OLEDs on Planarized Light Outcoupling-Enhancing Structures in Plastic. *Org Electron.* **111**, (2022).

62. Guo, F. et al. The Fabrication of Color-tunable Organic Light-emitting Diode Displays via Solution Processing. *Light Sci. Appl.* **6**, pagee17094 (2017).

63. Song, J., et al. Organic Light-Emitting Diodes: Pushing Toward the Limits and Beyond. *Adv. Mater.* **32**, 1907539 (2020).

64. Liang, J. et al. Origins and Influences of Metallic Lead in Perovskite Solar Cells. *Joule* **6**, 816–833 (2022).

65. Hu, J., et al. Organoammonium-Ion-based Perovskites Can Degrade to $Pb^0$ via Amine-Pb(II) Coordination. *ACS Energy Lett.* **6**, 2262–2267 (2021).

66. Lin, W. C. et al. In Situ XPS Investigation of the X-ray-triggered Decomposition of Perovskites in Ultrahigh Vacuum Condition. *Npj Mater. Degrad.* **5**, 13 (2021).

67. Birkhold, S. T. et al. Interplay of Mobile Ions and Injected Carriers Creates Recombination Centers in Metal Halide Perovskites under Bias. *ACS Energy Lett.* **3**, 1279–1286 (2018).

68. Yoo, E. et al. Bifunctional Resistive Switching Behavior in an Organo lead Halide Perovskite Based $Ag/CH_3NH_3PbI_{3-X}Cl_X$/FTO Structure. *J. Mater. Chem. C* **4**, 7824–7830 (2016).

69. John, R. A. et al. Reconfigurable Halide Perovskite Nanocrystal Memristors for Neuromorphic Computing. *Nat. Commun.* **13**, 2074 (2022).

70. Di, J. et al. Recent Advances in Resistive Random Access Memory Based on Lead Halide Perovskite. *InfoMat.* **3**, 293–315 (2021).